\documentclass[sigconf]{acmart}

\AtBeginDocument{%
  \providecommand\BibTeX{{%
    \normalfont B\kern-0.5em{\scshape i\kern-0.25em b}\kern-0.8em\TeX}}}

\copyrightyear{2022}
\acmYear{2022}
\setcopyright{rightsretained}
\acmConference[ISCA '22]{The 49th Annual International Symposium on Computer Architecture}{June 18--22, 2022}{New York, NY, USA}
\acmBooktitle{The 49th Annual International Symposium on Computer Architecture (ISCA '22), June 18--22, 2022, New York, NY, USA}
\acmDOI{10.1145/3470496.3527441}
\acmISBN{978-1-4503-8610-4/22/06}

\begin{document}

\title{Managing Reliability Bias in DNA Storage}


\author{Dehui Lin}
\affiliation{%
 \institution{Department of Computer Science}
 \country{National University of Singapore}}
\email{e0203126@u.nus.edu}

\author{Yasamin Tabatabaee}
\affiliation{%
 \institution{Department of Computer Science}
 \country{University of Illinois Urbana-Champaign}}
\email{syt3@illinois.edu}

\author{Yash Pote}
\affiliation{%
 \institution{Department of Computer Science}
 \country{National University of Singapore}}
\email{e0384154@u.nus.edu}

\author{Djordje Jevdjic}
\authornote{The corresponding author}
\affiliation{%
 \institution{Department of Computer Science}
 \country{National University of Singapore}}
\email{jevdjic@comp.nus.edu.sg}


\renewcommand{\shortauthors}{Lin et al.}

\begin{abstract}

DNA is emerging as an increasingly attractive medium for data storage due to a number of important and unique advantages it offers, most notably the unprecedented durability and density. While the technology is evolving rapidly, the prohibitive cost of reads and writes, the high frequency and the peculiar nature of errors occurring in DNA storage pose a significant challenge to its adoption.

In this work we make a novel observation that the probability of successful recovery of a given bit from any type of a DNA-based storage system highly depends on its physical location within the DNA molecule. In other words, when used as a storage medium, some parts of DNA molecules appear significantly more reliable than others. We show that large differences in reliability between different parts of DNA molecules lead to highly inefficient use of error-correction resources, and that commonly used techniques such as unequal error-correction cannot be used to bridge the reliability gap between different locations in the context of DNA storage. We then propose two approaches to address the problem. The first approach is general and applies to any types of data; it stripes the data and ECC codewords across DNA molecules in a particular fashion such that the effects of errors are spread out evenly across different codewords and molecules, effectively de-biasing the underlying storage medium and improving the resilience against losses of entire molecules. The second approach is application-specific, and seeks to leverage the underlying reliability bias by using application-aware mapping of data onto DNA molecules such that data that requires higher reliability is stored in more reliable locations, whereas data that needs lower reliability is stored in less reliable parts of DNA molecules. We show that the proposed data mapping can be used to achieve graceful degradation in the presence of high error rates, or to implement the concept of approximate storage in DNA. All proposed mechanisms are seamlessly integrated into the state-of-the art DNA storage pipeline at zero storage overhead, validated through wetlab experiments, and evaluated on end-to-end encrypted and compressed data.

\end{abstract}



\maketitle

\section{Introduction}
\label{sec:intro}

The digital universe keeps growing exponentially, mostly due to widespread  use and improved capabilities of image and video capturing devices~\cite{seagate:data, jevdjic:approximate}.  At the same time, the density and the production rate of conventional storage devices increase at much slower rates~\cite{seagate:data, ceze:molecular, bornholt:dna}, reducing our ability to preserve all the data we generate. 

This widening gap between the storage demand and supply can be bridged through a new and radical storage technology that uses DNA as a storage medium~\cite{church:next, goldman:towards} and offers a number of important and unique advantages: 
\begin{itemize}
\item Unparalleled Density. To illustrate, all the data stored in Facebook’s datacenter in Oregon, which is entirely dedicated to storage of high-density archived data, could fit into a sugar cube when stored in DNA, whereas our entire digital universe could fit into several bottles of DNA~\cite{bornholt:dna}.
\item Unmatched Durability. Depending on the method of preservation, data stored in the DNA format could last for hundreds of thousands of years~\cite{grass:robust}. This is in stark contrast to conventional storage technologies that retain data for a few years or decades, requiring new hardware acquisition and data transfer.
\item Eternally Relevant Interfaces. While the read/write interfaces of all storage devices eventually become obsolete, humans will always have an existential interest to read and write DNA~\cite{bornholt:dna}.
\item Efficient Random Access. Leveraging PCR, one of the fundamental reactions in biochemistry, allows us to selectively extract and read only the object of interest among petabytes of data~\cite{bornholt:dna, yazdi:portable, yazdi:rewritable,organick:random, tomek:driving, tomek:promiscuous}. The key implications are that both the cost and the latency of read operations are constant, regardless of the amount of data stored in the system. 
\item Efficient data manipulation. A number of operations, such as copying vast amounts of data or executing intelligent queries~\cite{stewart:content, bee:molecular}, can be conveniently performed in the molecular domain. 
\end{itemize}

While the technology is evolving rapidly, and the first fully automated end-to-end DNA storage system has recently been demonstrated~\cite{takahashi:demonstration}, a number of major challenges remain to be overcome. The primary obstacle to DNA storage adoption is the prohibitive cost of reads and writes. Writing digital data into DNA (synthesis) and converting it back into digital form (sequencing) incurs prohibitive capital and operating costs~\cite{ceze:molecular}, impeding not only the commercial deployment of DNA storage, but also the research efforts. The problem is further exacerbated by high rates and complex nature of errors in DNA. Namely, the processes involved in DNA synthesis (write), data access, and DNA sequencing (read), and even the software processing of DNA sequences are approximate in nature and highly prone to errors~\cite{goldman:towards, organick:random,bornholt:dna}. The lower the cost of these processes, the higher the error rates~\cite{organick:random,takahashi:demonstration}. As a result, significant amounts of redundant  resources must be invested to allow for full recovery of binary data from DNA molecules. As such, efficient handling of errors in DNA storage is critical to reducing the cost~\cite{yazdi:portable, bornholt:dna, grass:robust, organick:random}.

In this work we make an important observation that, from the system point of view, some parts of DNA molecules represent significantly more reliable locations to store data compared to other parts. We find that the relative order of reliability of different locations within a strand can be easily established at the time of encoding, while the magnitude of the skew depends, in a complex manner, on a few parameters explained later. Interestingly, the reason behind this skew is not directly related to the underlying chemistry, but to the fact that each piece of data must be reconstructed from multiple noisy copies that suffered from insertions and deletions of DNA bases; finding the consensus among such noisy copies requires an algorithmic step whose ability to correctly reconstruct the original bits fundamentally depends on the position of the bits within the molecule. As long as there are any insertion/deletion errors present in the process of reading and writing, all DNA storage architectures will observe reliability skew across positions within each molecule. If the storage system is not aware of it,  the existence of the skew will lead to considerable inefficiencies in provisioning of costly error-correction resources, and will lead to higher costs for read and write operations.

Thinking of DNA storage as an information channel, one could theoretically compensate for the skew in reliability by employing uneven error correction, which would be a natural solution to a biased channel. Under uneven error correction, the ECC redundancy is provisioned to each data location in proportion to its reliability properties; less reliable locations receive proportionately more error-correction resources, and vice versa, potentially reversing the skew~\cite{guo:approximate, jevdjic:approximate}. However, to optimally provision redundancy, we need to know the exact magnitude of the reliability skew, which in turn requires the knowledge of the precise error profile of all chemical processes involved. Given that the durability of DNA is measured in thousands of years and that DNA reading (sequencing) technologies evolve rapidly, optimally configuring unequal redundancy for its use in 1000 years is impossible. In fact, DNA storage is such a dynamic and unpredictable stochastic channel that even with the full knowledge of the synthesis and sequencing technologies, it is nearly impossible to accurately estimate the magnitude of the reliability skew.

To effectively address this reliability bias, we propose two techniques. The first technique, called $Gini$, \footnote{Inspired by the Gini wealth inequality index.}, removes any positional reliability bias in DNA storage. Gini amortizes the impact of errors by interleaving error-correction codewords across many DNA molecules such that the impact of errors is evenly spread across a large group of ECC codewords.

Our second proposed technique, called $DnaMapper$, builds on the insight that different pieces of $data$ may have different reliability needs~\cite{sampson:approximate, guo:approximate, jevdjic:approximate}. Having a storage system that offers different classes of reliability on  one hand (the set of storage locations that occupy a given position in each molecule being a reliability class), and data with diverse reliability needs on the other hand, DnaMapper performs an application-aware mapping of data onto DNA molecules such that data requiring higher reliability is stored in more reliable locations, whereas the data that needs lower reliability is stored in less reliable parts of DNA molecules.

To summarize , this work makes the following contributions:
\begin{itemize}
\item We are the first to make the observation that all DNA storage architectures experience reliability skew, such that some parts of the molecules are significantly more reliable than others and the relative order of reliability of different locations within a molecule can be easily and statically determined.

\item We propose Gini, a technique that spreads the impact of errors in DNA storage evenly across ECC codewords, such that every ECC codeword is affected by a nearly identical number of errors. As a result, every error has a similar probability of being corrected, regardless of its spatial origin. In contrast to conventional techniques like uneven error-correction~\cite{guo:approximate, jevdjic:approximate}, Gini is guaranteed to always evenly spread the errors, regardless of the underlying error profile and sequencing technology used. Gini is applicable to any type of data and can be used to improve the reliability of the storage system, and/or to reduce the amount of expensive reads and writes. 

\item We propose DnaMapper, a priority-based mapping scheme that maps data onto DNA according to the reliability needs of different bits, such that the data that requires high reliability is stored in most reliable parts of DNA molecules, and conversely, data whose corruption is more tolerable is stored in less reliable locations. This mapping scheme is general and can be applied to any type of data that has a notion of quality and is ideally suited for images and videos. 
\end{itemize}

We evaluate DnaMapper using a simple and effective-enough heuristic for determining the relative order of bits in standard JPEG images according to their reliability needs. While other heuristics would most likely yield better results~\cite{guo:approximate, jevdjic:approximate}, this method is much simpler and does not need to maintain any metadata to describe the mapping and therefore imposes no storage overhead. This method is also content-agnostic, which allows for approximate storage of end-to-end encrypted data, unlike previous approximate storage methods~\cite{guo:approximate, jevdjic:approximate}.

 All proposed mechanisms involve no storage overhead and can be easily integrated into any DNA storage pipeline. In fact, we integrate both Gini and DnaMapper into the state-of-the-art DNA storage pipeline~\cite{organick:random} through simple data reshuffling. We also showcase the feasibility and practicality of the proposed techniques on a tiny scale in the wetlab, where we successfully retrieved from DNA and decoded all images stored in all proposed formats. Using simulation, we show that DnaMapper can provide graceful degradation in case of higher-than expected error rates, as well as reduce the reading cost by up to 50\% while retrieving the images of the same quality as the baseline system. Gini reduces the reading cost up to 30\% and writing cost up to 12.5\%.

The rest of the paper is organized as follows. Section~\ref{sec:background} provides background on DNA-based storage architectures and prevailing error-correction mechanisms used. Section~\ref{sec:consensus} introduces and analyzes the problem of consensus finding and provides the intuition behind the reliability skew. Sections~\ref{sec:gini} and~\ref{sec:dnamapper} describe the proposed mechanisms.  Section~\ref{sec:methodology} describes our experimental methodology, while Section~\ref{sec:evaluation} presents the evaluation highlights. We conclude in Section~\ref{sec:conclusion}.

\section{Background}
\label{sec:background}
\subsection{Storage and Retrieval of Data in DNA}
Writing data into DNA relies on DNA synthesis, which is the chemical process of creating artificial DNA molecules.
Unlike biological DNA molecules, which contain billions of base pairs (bp), artificial DNA molecules ($strands$) are limited in length due to the practical limitations of artificial DNA synthesis. Today's cost-efficient synthesis creates strands which are only 100-200 bp long and can hold less than 50 bytes of data. Large files therefore must be split into smaller chunks~\cite{bornholt:dna, organick:random}, and to be able to reassemble the original data from the chunks, each chunk must contain the ordering information ($index$).

Chunks of binary data can be encoded into a sequence of \{A, C, G, T\} bases using a variety of coding techniques. Some coding techniques
seek to avoid immediate repetition of bases, also known as  $homopolymers$ (e.g., AAA)~\cite{bornholt:dna} to reduce the chance of sequencing errors on some machines, while others seek to balance the ratio of G+C bases versus A+T (aka $GC-content$) to improve the chances of successful synthesis~\cite{yazdi:rewritable, yazdi:portable}.  Although these constraints come at the expense of coding efficiency, they can also be useful in detecting and correcting errors that cause the imposed constraints to be violated~\cite{yazdi:rewritable}. Without loss of generality, in this work we assume a simple coding scheme in which two bits of data are directly mapped to one DNA base (e.g., 00 = A, 01 = C, 10 = G, 11 = T), which achieves the maximum information density. 

Once encoded into DNA strings, chunks belonging to different files are tagged with different special sequences, known as $primers$, with one
primer prepended at the beginning and the other appended at the end of each molecule. All chunks belonging to the same file are tagged with 
the same pair of primers. The primers are used as parameters of the PCR reaction, which essentially provides a chemical data lookup mechanism.
Each pair of primers logically represents a key in a key-value store~\cite{bornholt:dna} and enables random access within large volumes of 
data~\cite{organick:random, yazdi:rewritable}. The final DNA  strings are synthesized into molecules, usually with millions of copies of each.

The first big step in retrieval of a file includes isolating the molecules with the correct primer pair through the process of selective amplification (PCR reaction). The isolated molecules are read using one of many available sequencing methods, which have different accuracy, throughput, and cost characteristics. The outcome of the sequencing process is a large collection of DNA strands ($reads$). The average number of reads per original molecule represents an important metric called sequencing $coverage$. A higher coverage implies higher chances of correctly reconstructing the original strands. Unfortunately, the sequencing coverage is directly proportional to the sequencing costs. Therefore, \textbf{minimizing the required sequencing coverage is crucial to reducing the cost of reading from DNA}. 

After filtering out the reads with incorrect primers, the remaining reads must be clustered~\cite{rashtchian:clustering} based on similarity, so that reads corresponding to different data chunks can be disambiguated. The similarity metric of interest is usually assumed to be \textit{edit distance}~\cite{organick:random, rashtchian:clustering}, which corresponds to the minimum number of operations (insertions, deletions, substitutions) required to convert one string into another. With each cluster containing a number of noisy copies of a single chunk of the original data, the next big step is to find the consensus among the noisy copies and produce the best estimate as to what the original strand was. This important step is the subject of the next section. After identifying the most probable original strand for each data chunk (cluster), the next step is to convert each DNA strand back into the binary form and reassemble the chunks into a single file using the ordering information encoded in every chunk. Error correction can be performed either before~\cite{yazdi:rewritable, yazdi:portable},  after~\cite{organick:random}, or during~\cite{bornholt:dna, goldman:towards} the reassembly, with the prevailing approach described below.

\subsection{Error Correction in DNA Storage}

\begin{figure}
	\includegraphics[width=1\columnwidth]{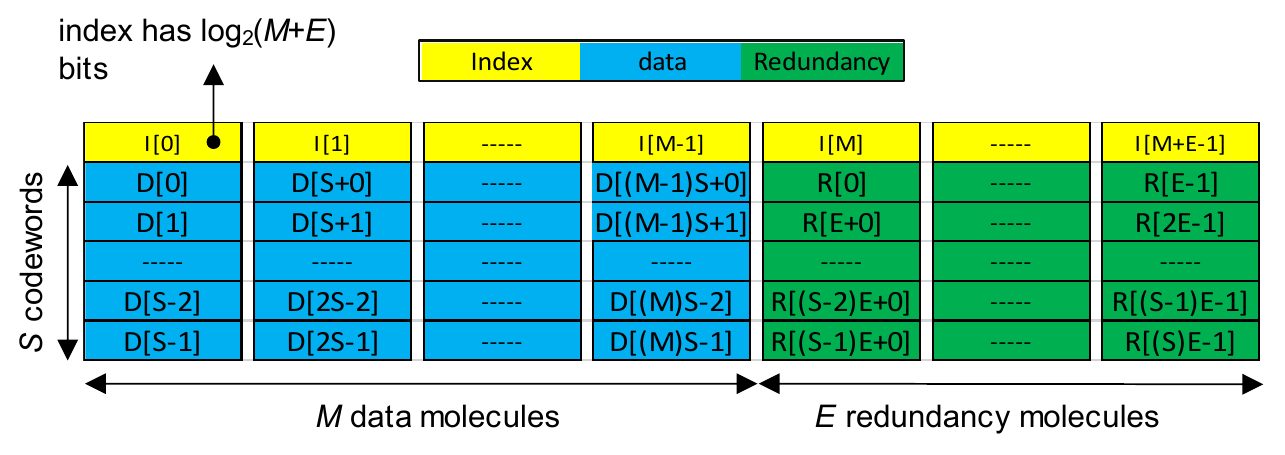}
	\caption{DNA Storage Architecture with Reed-Solomon ECC~\cite{organick:random}}
	\label{fig:error_correction}
\end{figure}

Figure~\ref{fig:error_correction} shows the state-of-the-art DNA storage architecture~\cite{organick:random} with Reed-Solomon Error Correcting codes. In this architecture, data is laid out into a matrix, with every DNA molecule representing a column in the matrix, while each row is a Reed-Solomon codeword. The whole matrix represents a unit of encoding/decoding. The key parameter of Reed-Solomon codes is the symbol size: for a symbol size of $n$ bits, the codeword must have ${2}^{n}$-1 symbols (the entire codeword thus has $n$(${2}^{n}$-1) bits). Based on the desired error correction capabilities, the symbols in a codeword are split between $M$ data symbols and $E$ redundant symbols, with $M$+$E$=${2}^{n}$-1; In each encoded unit there are $M$ data molecules and $E$ redundancy molecules for error correction. As explained in the previous subsection, each molecule must have an index required for ordering. Note that the redundant symbols are synthesized into separate DNA molecules and therefore, they need to be ordered as well, so that they can be placed in the correct column within the matrix at the decoding time. The consequence of this is that the ordering information cannot be protected by error correction. The ordering index must contain at least $\log_{2}(M+E)$ bits, which is ($\log_{2}(M+E)$)/2 bases. Given that $M$+$E$=${2}^{n}$-1, the index will have $n$ bits or $n$/2 nucleotide bases, which equals the size of the symbol. In the example in Figure~\ref{fig:error_correction}, the indexes are placed at the beginning of each molecule, while the data symbols are mapped column-wise, molecule by molecule. 

A Reed-Solomon codeword with $E$ redundant symbols can correct up to $E$ erasures within a codeword, or detect and correct up to $E/2$ errors. Erasures are types of errors which happen when data is missing/wrong but the location of the missing/wrong data is known. Erasures generally are less costly because they do not require detection, however in this architecture they usually happen in case of clustering errors or loss of entire clusters; when a cluster at index $i$ is missing, an erasure is present in every row of the matrix at location $i$. It is important to note that in this architecture, errors such as insertions and deletions within each molecule (column) are detected as substitution errors in the corresponding codeword (row) and corrected as such.  

Note that the data layout Figure~\ref{fig:error_correction} is similar to the data layout in a RAID disk setup~\cite{raid}. In our case a column represents a DNA molecule, whereas in conventional storage a column would represent a sector. A sector failure would constitute an erasure, which in the DNA storage context maps to a failure to read a particular DNA strand. 

\section{Consensus Finding and Reliability Skew}
\label{sec:consensus}

\begin{figure*}
	\includegraphics[width=1\textwidth]{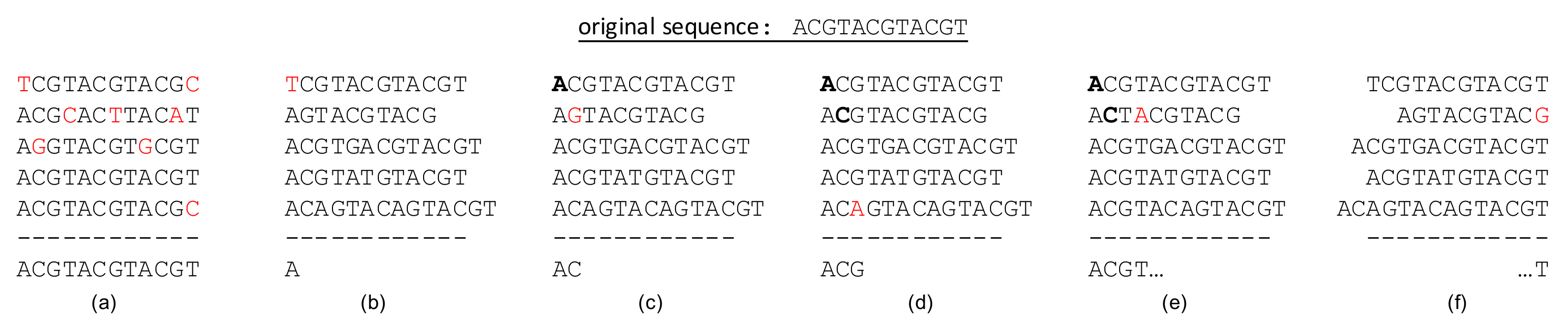}
	\caption{Simplified example of finding a consensus string}
	\label{fig:example}
\end{figure*}
After clustering DNA reads based on edit distance, each cluster is separately processed to find the most likely original DNA sequence corresponding to the noisy reads within that cluster. We can formulate this problem as follows: Let $s$ be an unknown string of length $L$ over alphabet $\Sigma$=\{A, C, T, G\}. We are given $N$ noisy copies of $s$, each generated independently by distorting $s$ at each position with probability $p$, where $N$ corresponds to the sequencing coverage. In other words, for each position $i \in [1..L]$, we either delete the $i$-th character of $s$ (deletion), or insert a character chosen uniformly at random from $\Sigma$ at that position (insertion), or replace the $i$-th character of $s$ with another character from the alphabet  (substitution), or keep it unchanged.  For the sake of simplicity, we assume that each of the error types occurs with probability $p/3$, but our model can be easily generalized to support different probabilities for each error type. Given these $N$ noisy copies, the goal is to reconstruct $s$. More formally, the goal is to find a string of length $L$ such that the sum of edit distances to all inputs is minimal across all strings of length $L$.  A string that minimizes the sum of edit distances to given inputs across all strings (not constrained to a given length) is known as \textit{edit distance median}. The problem of finding the edit distance median is shown to be NP-complete~\cite{nicolas:hardness}, and so is our problem of finding the median of length $L$, which we refer to as \textit{constrained edit distance median}.

When the input strings originate from the same original string, the task belongs to a class of problems in information theory which are commonly referred to as \textit{trace reconstruction} problems. The study of a variation of this problem was initiated by Batu et al.~\cite{batu:reconstructing}, considering only binary deletion channels. They proposed a simple algorithm called Bitwise Majority Alignment and showed that it can be used to exactly reconstruct the original string with high probability using $O(\log{}L)$ noisy copies when the error rate is $O(1/\log{}L)$. A subsequent work~\cite{kannan:more} extended this result to binary channels with both insertions and deletions for an error rate of $O(1/\log{}^2L)$. Viswanathan and Swaminathan~\cite{viswanathan:improved} improved this result by presenting an algorithm that needed the same number of traces for insertion and deletion probabilities of $O(1/\log{}L)$. Several researchers have advanced these results further as explained in a recent survey~\cite{bhardwaj:trace}, proposing lower and upper bounds for the number of traces necessary and sufficient for reconstruction with high probability~\cite{krishnamurthy:trace, duda:fundamental}. It is important to note that while all these important theoretical findings established the relationship between the length of the strings ($L$), error rate ($p$), and the probability of successful exact reconstruction of the entire string, they have not looked at the probability of reconstruction of individual characters within the string and how it relates to the character positions in the string.

\subsection{Reliability Skew}
To give an intuition as to why the reliability skew exists and why the length of the original strand poses a challenge to the reconstruction, let's consider an example in Figure~\ref{fig:example} with five noisy copies of an original sequence. When the noisy copies contain only substitutions (Figure~\ref{fig:example}a), we can perform a simple majority vote for each position independently and correctly reconstruct the original string even when the coverage (number of inputs) is small and the error rate is high. However, if we allow insertions and deletions to happen (Figure~\ref{fig:example}b), we cannot apply such a simple reconstruction procedure. First, the copies can have different lengths. Second, the characters may not be in their original positions even if all strings were of the same, desired length. To perform the reconstruction, we must place the characters in each string into their original positions. Because the characters at the beginning and the end cannot be misplaced by too many positions, it is best to start fixing the misplaced characters from each end of the sequence.

Looking at the first column in Figure~\ref{fig:example}b, we see that A is the most likely character and T is an outlier in the first string. We can safely assume that the first character of the original sequence is A, but to continue using the first string we must understand what kind of a distortion it had suffered and try to undo it to guarantee the correct placement of characters. Because the second and third characters are CG in most of the sequences, including the first one, we can assume that this was most likely a case of substitution. So we correct the first string by converting T to A and move forward by one position (Figure~\ref{fig:example}c).  

\begin{figure}
	\includegraphics[width=1\columnwidth]{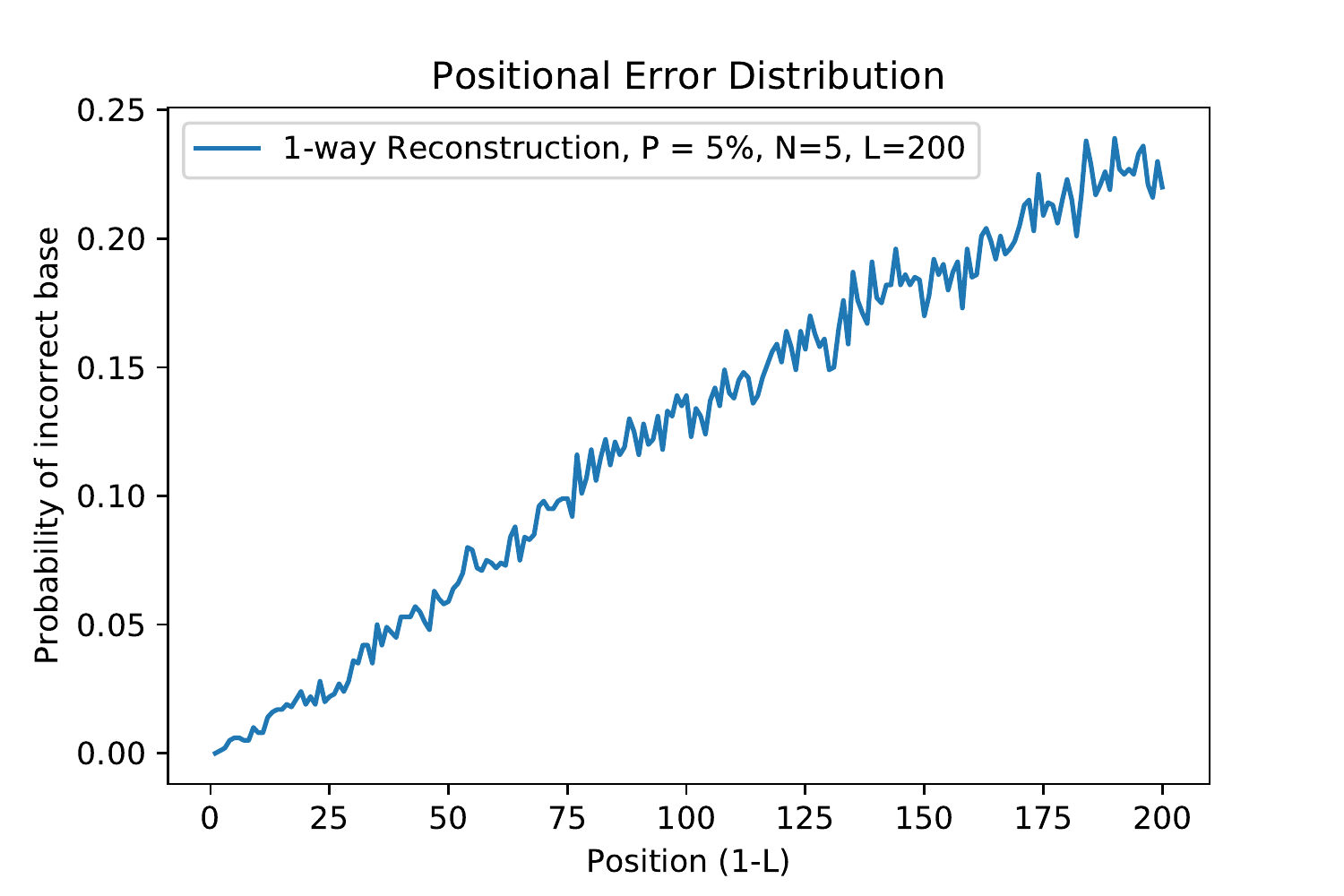}
	\caption{Probability of correctly identifying a base in the string based on its position.}
	\label{fig:nondp-1way}
\end{figure}

\begin{figure}
	\includegraphics[width=1\columnwidth]{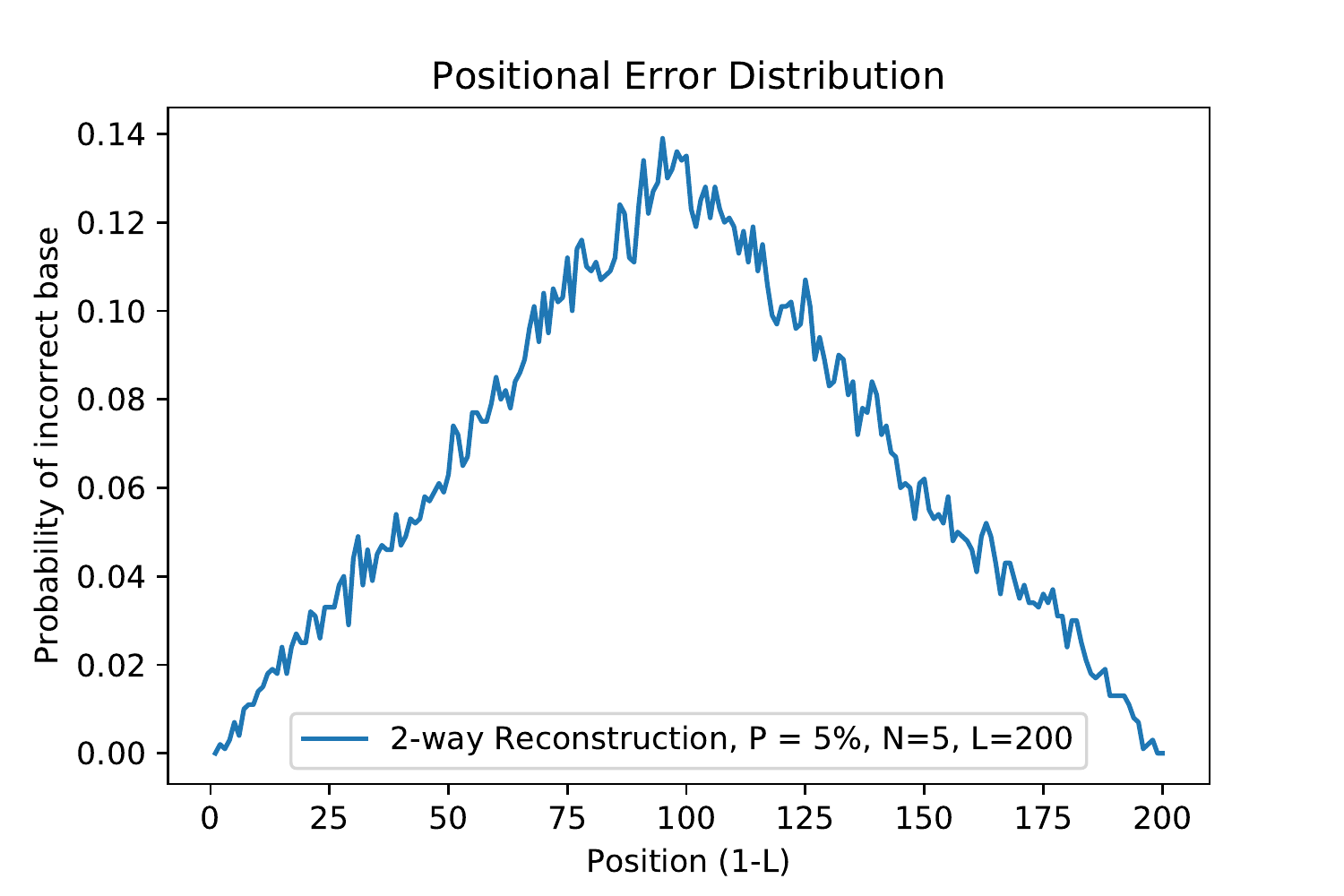}
	\caption{Probability of correctly identifying a base in the string based on its position in case of a 2-way reconstruction.}
	\label{fig:nondp}
\end{figure}

Looking at the second column, we can conclude that the consensus character is C (Figure~\ref{fig:example}c), with G being an outlier in the second string. Observing that the next two characters in all strings are GT except for the second one (TA), we can conclude that the second string likely suffered from a deletion of C. We therefore fix the second string by inserting C before G and move on to the next position (Figure~\ref{fig:example}d). In the third column, we can see that the consensus character is G with A being an outlier in the last string. To fix the error, we look at the next two characters, which are GT for the last string and TA for all other strings. We can assume that A was likely inserted before G in the last string. We fix the error and move on (Figure~\ref{fig:example}e). 
 
This example illustrates the key problem with consensus finding algorithms: whenever we encounter an outlier, we have to make an assumption as to what the error was and then correct the error based on that assumption. If our assumption is not correct, the original error plus any error introduced by our mis-correction will propagate to the next position. As we advance towards the end of the sequence, the errors accumulate and our ability to find a consensus diminishes. This phenomenon is illustrated in Figure~\ref{fig:nondp-1way}, which shows the probability of correctly identifying a nucleotide base based on its position within the string. As we can see, the error increases sharply as the base index increases. An obvious implication is that longer strings will have higher maximum error probability. The problem illustrated in Figure~\ref{fig:nondp-1way} is a direct result of the fact that placing a base in its correct position cannot be done independently of other bases due to deletions and insertions, which was not the case in Figure~\ref{fig:example}a, which assumes that only substitutions can happen. 

Fortunately, the problem of consensus finding for linear structures is symmetric; we can align all the strings to the right and start the reconstruction process from the other end, as illustrated in Figure~\ref{fig:example}f. In this case, we can use only the first (i.e., better) half of the string reconstructed from left to right, and the other half from the string reconstructed from right to left to create the final consensus string as the best of both worlds. Figure~\ref{fig:nondp} illustrates the probability of correctly identifying a base based on its position after applying the described 2-way reconstruction procedure. As we can see, the  probability of error is low at at the ends of the string, then gradually increases towards the middle, and is the highest in the middle of the string.

\begin{figure}
	\includegraphics[width=1\columnwidth]{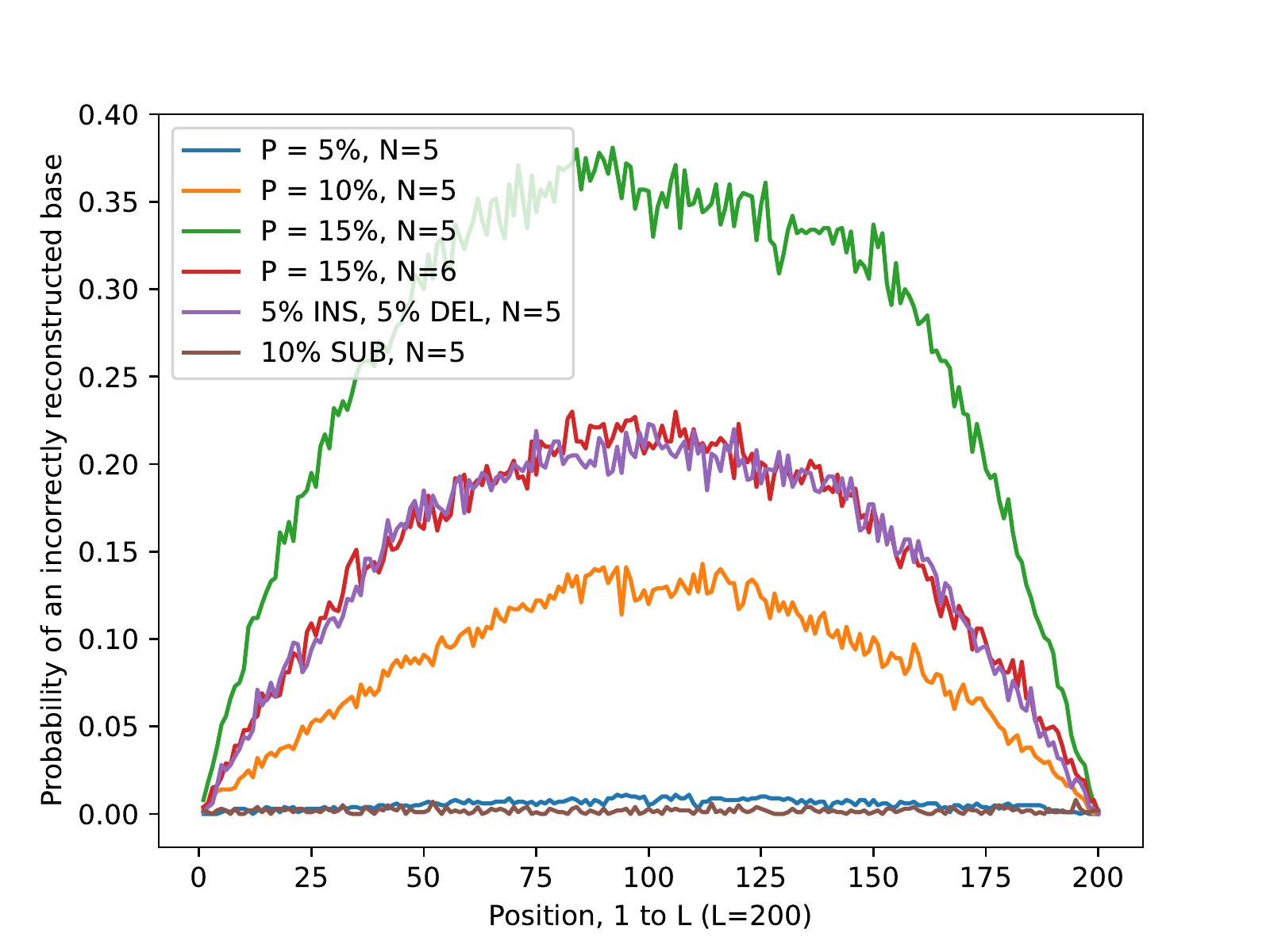}
	\caption{Reliability skew observed in the state-of-the-art  trace reconstruction algorithm~\cite{sabary:reconstruction}.}
	\label{fig:skewsabary}
\end{figure}

The consensus finding approach explained above and different variations of it are commonly used in DNA data storage pipelines ~\cite{organick:random}. However, recent work on trace reconstruction has introduced other practical algorithms with various definitions and optimization criteria. Most notably, Sabary et al ~\cite{sabary:reconstruction} have proposed an iterative reconstruction algorithm that solves the \textit{DNA Reconstruction Problem} with the goal of finding an estimation with minimal edit distance from the original string. This algorithm outperforms other algorithms used in practice in terms accuracy~\cite{sabary:reconstruction}. Although this algorithm does not follow the two-sided approach explained above, the skew is still present for various parameters as shown in Figure~\ref{fig:skewsabary}, except in the case where only substitutions are present (purple line). The shape of the skew is the same, with the peak being higher for smaller $N$ and/or larger $P$.\footnote{The experiments were performed using the available code~\ref{fig:skewsabary}. The algorithm occasionally produces the result of incorrect length, and we have excluded such strands while plotting Figure~\ref{fig:skewsabary}.}  

In Figure~\ref{fig:skewsabary}, the top four lines (as per the legend) assume equal distribution of substitutions, deletions, and insertions (1/3rd each). For the last two lines, purple and brown, we change the breakdown of the three error types. Note that in case of the brown line (10\% substitutions, no indels), there is no observable skew and the algorithm easily reconstructs the data. This is expected, as substitutions alone, similarly to bit-flips, do not create any skew. In contrast, the orange line with the same error rate of 10\% but with equal representation of the error types, shows significant skew. In fact, even a 2x lower error rate (5\%) with equal representation of the error types (blue line) shows some skew in the middle and presents a bigger challenge for reconstruction compared to 10\% substitution errors (brown).

Let's now compare the green and purple lines, which contain the same number of insertions and deletions (indels), while the green line further has 5\% substitution errors. While a substitution-only error rate of 10\% makes no impact on the skew (brown curve), the addition of just 5\% substitution errors has a significant impact the presence of indels, as measured by the difference between purple and green lines. In conclusion, while in isolation substitutions do not create any skew and are easy to correct, they amplify the skew and complicate the reconstruction in the presence of indels. Interestingly, adding an extra strand (red line) has a similarly strong impact as removing substitutions (purple line).

\subsection{Fundamental Nature of the Reliability Skew}
A natural question to ask is whether the observed skew is solely a result of using a particular algorithm or is it an unavoidable property of trace reconstruction in the presence of insertions/deletions. To determine the answer empirically, we experiment with strings of small length so that optimal trace reconstruction is possible in reasonable time through a brute force search for an edit distance median of the target length. If we can observe the error peak in the output of optimal algorithms, it can be seen as evidence towards the assertion that the skew is a part of all optimal reconstruction algorithms. 

Without loss of generality, we assume a binary alphabet (\{0, 1\}, instead of \{A, C, G, T\}). For practical reasons, we limit the length of the original bit string to 20 bits. For a large number of input bit strings of length 20, we construct a noisy cluster of size $N$ through insertions, deletions, and substitutions with total probability of $p$=20\%. We then find all strings of length L such that the sum of edit distances to all strings in the cluster is minimal. Importantly, if there are multiple such strings, we pick one in an adversarial manner, such that the selected string is more accurate towards the middle than towards the ends (compared to the original string), in an effort to create a skew opposite from the one we expect to see.  
\begin{figure}
	\includegraphics[width=1\columnwidth]{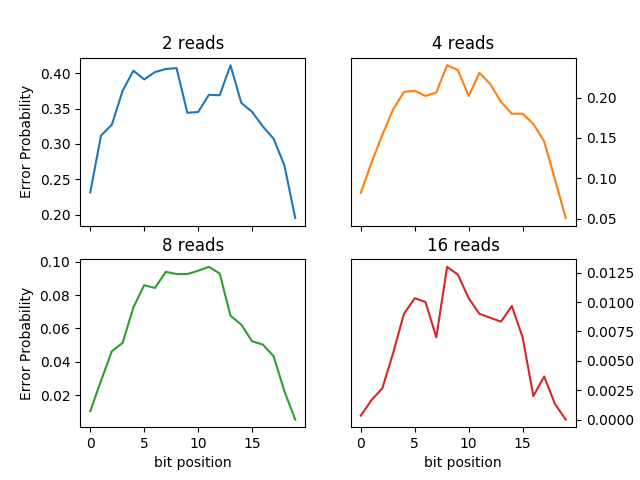}
	\caption{The probability of incorrectly reconstructing a bit as a function of its position in a bit string of length L=20 for N=2, 4, 8 and 16 and probability of error p=20\%.}
	\label{fig:fundamental}
\end{figure}

Figure~\ref{fig:fundamental} shows the probability of incorrectly reconstructing a bit as a function of its position in a bit string of length L=20 for N=2, 4, 8 and 16 and probability of error p=20\%. Interestingly, while the higher number of reads reduces the peak error probability, the shape of the curve doesn't change singificanlty. We can see that even when an optimal algorithm uses all available degrees of freedom to reverse the expected bias, the reliability skew is still present and significant.

\subsection{Implications on Reliability and Future Trends}
The shape of the error probability as a function of base position curve has profound implications on reliability. The bases at the beginning and the end of each molecule present reliable places to store data, whereas the bases in the middle area are significantly less reliable.  The trends in DNA sequencing (reading) and synthesis (writing) suggest that the skew in reliability between different positions will have even more significant consequences in the future. First, the synthesis process improves over time producing longer molecules~\cite{ceze:molecular}. Longer molecules are desirable because they result in proportionately fewer data chunks per file, reducing the overheads of primers and ordering indexes. However, longer molecules make the problem of consensus finding more challenging and significantly exacerbate the problem of the reliability skew. Second, new sequencing technologies using nanopores dramatically reduce the reading costs~\cite{jain:nanopore}, but introduce much higher (an order of magnitude) error rates, significantly complicating the consensus finding step~\cite{yazdi:portable, duda:fundamental} and resulting in even steeper error probability curves. Finally, lower sequencing coverage is desirable as it directly reduces the cost of sequencing~\cite{organick:random}. However, lower coverage implies significantly harder and less accurate consensus finding.  All these trends suggest that the reliability bias is likely to increase in the future, and DNA-based storage systems must be aware of the inherent reliability skew to avoid significant over-provisioning of error-correcting resources.

\section{Flattening the Curve}
\label{sec:gini}

\begin{figure}
	\centering
	\includegraphics[width=1\columnwidth]{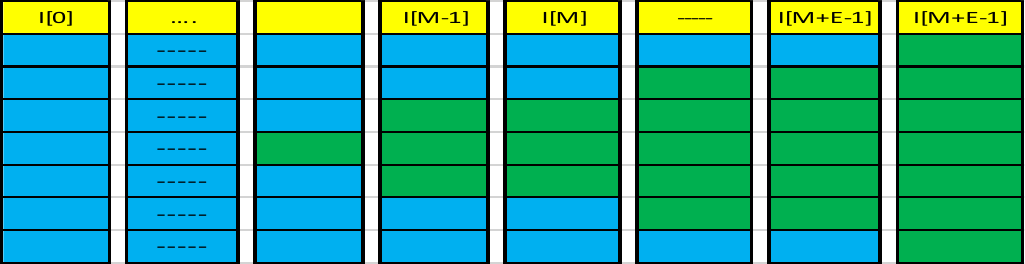}
	\caption{Example of an architecture with uneven provisioning of redundancy}
	\label{fig:unequal}
\end{figure} 

\subsection{Unequal Error Correction}
Given the reliability bias in DNA storage, one may be tempted to apply unequal error correction, given that DNA storage allows for high-precision ECC tuning. Although the complex error-correction mechanism depicted in Figure~\ref{fig:error_correction} requires an entire matrix to be encoded as a unit, it is still possible to provide a custom amount of redundancy for each codeword (row) within the matrix, with high precision. Figure~\ref{fig:unequal} demonstrates what an uneven ECC would look like, where the most reliable locations in all molecules (the first and the last row) receive the least amount of redundancy, while the rows in the middle receive significantly more redundancy.  In this case there is no more clear separation between data-only and redundancy-only molecules. While most of the molecules still contain only data symbols, and some molecules may still contain only redundancy symbols,  there are a number of molecules that contain a mix of data and redundancy.

While redundancy can be provisioned in every row in a very precise manner, there is no way to know how much redundancy each row should receive in advance. The desired variability in redundancy may change when the sequencing method changes, or even when the coverage changes. As shown in Figure~\ref{fig:skewsabary}, increasing the sequencing coverage from 5 to 6 may change the magnitude of the skew by 2x, and per-strand coverage is not possible to control~\cite{organick:random}. Yet, to implement unequal redundancy, we would have to assume a particular skew curve and fix the redundancy in each row at the time of encoding, which clearly is not a solution that can stand the test of time, given that DNA is a durable, archival storage medium that lasts for thousands of years~\cite{grass:robust} and the sequencing methods are more than likely to change multiple times during the lifetime of data.

Furthermore, even if we had the perfect knowledge of the sequencing technology and the protocol to be used at the time of reading, and even if we somehow knew the target sequencing coverage and the exact algorithm to be used for consensus finding, even in this case the unequal redundancy approach has serious problems because coverage is never fixed across all clusters. Instead, coverage follows the Gamma distribution~\cite{organick:random}, with a significant variation in size across individual individual cluster. As such, despite having the desired average coverage, some clusters will have more reads and some fewer, and the level of skew in individual clusters will be different, and very few of them will have the exact average coverage that we statically provisioned the skew for at the time of encoding.

\subsection{Gini} 
The baseline architecture~\cite{organick:random} depicted in Figure~\ref{fig:error_correction} provides great protection against erasures, i.e., the losses of entire molecules during sequencing. In case of erasures, a single substitution error is seen in every codeword. However, ordinary deletions and insertions will accumulate a great deal of error in the middle of each molecule (column). As a result, the rows in the middle, each of which maps to an ECC codeword, will suffer many more errors compared to the codewords towards the ends. 

\begin{figure}
	\centering
	\includegraphics[width=1\columnwidth]{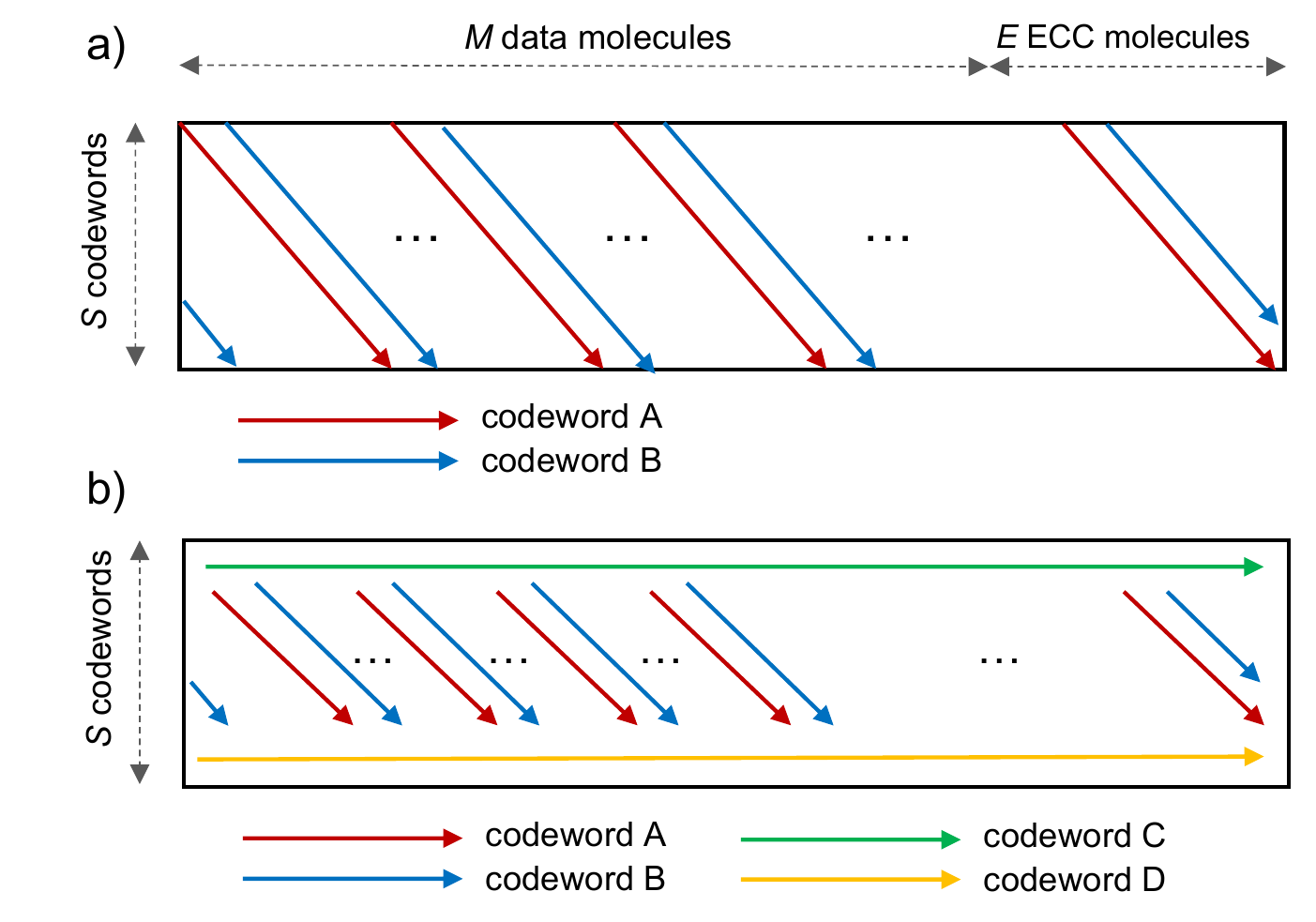}
	\caption{Codeword interleaving in Gini}
	\label{fig:gini_design}
\end{figure}

To propose an effective solution, we take an inspiration from the coding theory in mobile communications, where a similar problem of positional error bias across both rows and columns of the encoding matrix has been observed~\cite{diagonal}. We make the observation that, while we cannot control the spatial distribution of errors, we can control the impact of such errors by carefully defining our codewords. Instead of distributing codewords across rows of the encoding matrix, we can distribute them diagonally, as shown in Figure~\ref{fig:gini_design}a. Since the number of columns is usually orders of magnitude larger than the number of rows, each codeword will wrap around the matrix many times, evenly cycling through all positions in all molecules. Consequently, the errors in the middle will be equally distributed across all codewords, unlike in the baseline where all errors coming from the middle of every molecule are concentrated in the same codeword. 

Gini removes all positional reliability bias in DNA storage by spreading the impact of errors evenly across a large group of ECC codewords. When it comes to erasures, Gini matches the capabilities of the baseline, as every symbol in every molecule belongs to a different codeword. Note that for this to happen, we must ensure that when wrapping a diagonal codeword around the matrix, we continue from the next column, as shown in Figure~\ref{fig:gini_design}a. Also note that we can decide to exclude arbitrary rows from this interleaving. For example, we could exclude the first and last rows and reserve them for very important data and treat them as separate codewords, while the rest of the codewords can be interleaved across the rest of the rows, as shown Figure~\ref{fig:gini_design}b, where we essentially created two reliability classes.

Gini does not change the number of errors that take place, but simply redistributes them in a way that equally impacts every codeword, allowing all errors to have a similar probability of successful correction, regardless of their spatial origin. Gini can be used to improve the reliability of the system and reduce the number of copies of each molecule that must be read by a sequencer, leading to commensurate savings in the reading cost. Gini can also be used to reduce the amount of error-correction resources while keeping the reliability constant, leading to savings in both DNA reads and writes. Gini is applicable to any type of data, requires no storage overhead and can be easily integrated into state-of-the-art DNA storage architectures.

\section{Application-Aware Data Mapping}
\label{sec:dnamapper}
 \begin{figure}
	\includegraphics[width=1\columnwidth]{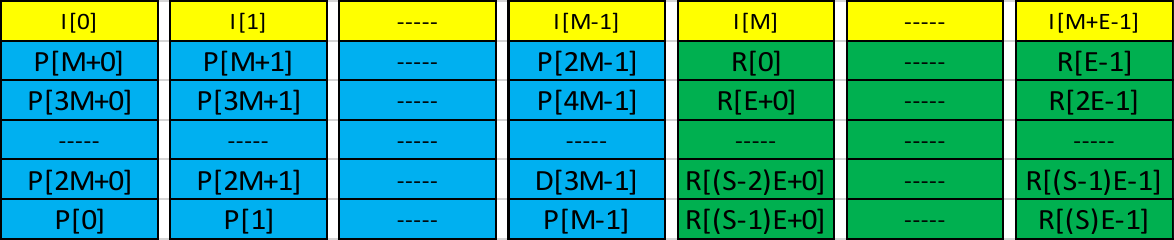}
	\caption{Priority-based mapping of data onto Reed-Solomon symbols with DnaMapper}
	\label{fig:dnamapper}
\end{figure} 
In this section, we describe how we can leverage the skew reliability across different parts of DNA molecules on the one hand, and the skew in reliability needs across different data bits on the other, to produce an optimal mapping of data onto molecules without a need to remove the bias in the storage medium.

\subsection{General Framework}
Given $N$ data bits of known reliability needs, and given $N$ storage cells of known reliability properties, what is the
optimal mapping of bits into cells that maximizes the retrieved data quality? Interestingly, the answer to this question does
not depend on any absolute values indicating reliability needs of bits, or any absolute value indicating reliability properties
of storage cells. It also does not matter whether cell A is 3x more reliable than cell B, or only by 2x. It is easy to show
(proof by contrapositive) that the optimal mapping will always be the one in which the bit with the highest reliability needs
is mapped to the cell of highest reliability, the bit with next highest needs is mapped to the cell of next highest
reliability, and so on. In other words, it is only the ranking of cells by reliability and the ranking of bits by reliability
needs that matters. This is of crucial importance in our context, because unlike the amplitude of the skew, the ranking of bases by reliability in DNA storage can
be statically determined and does not depend on the retrieval process.

\subsection{Mapping data onto DNA molecules}
As described in Section~\ref{sec:background}, each file is split into chunks of fixed size and each chunk is mapped into a separate molecule. Given $M$+$E$ molecules, $\log_{2}(M+E)$ bits (which is ($\log_{2}(M+E)$)/2 bases) must be reserved within each molecule for the ordering information (index), and the data bits are mapped into remaining positions.  

\subsubsection{Baseline Mapping}
The baseline mapping of data onto molecules is shown in Figure~\ref{fig:error_correction}. We place the first chunk of data into molecule 0, the next chunk of data into molecule 1, etc. The last chunk of data is placed into molecule $M$-1, and the remaining $E$ molecules are redundant.

\subsubsection{Priority-Based Mapping}
Recall that the bases at the beginning and the end of DNA molecules represent reliable data locations, whereas the bases in the middle are unreliable. Due to symmetry, each position in the molecule has a corresponding position of the same reliability. Because the ordering information is of utmost importance, we place it at the most reliable part of each molecule, which is the first (or last)  location. Note that the index placement in this scheme happens to be the same as in the baseline mapping. The next most reliable locations are the last bases of each molecule. We therefore strip 2$M$ most important data bits across $M$ molecules, placing them in the next most reliable set of bases, which is the last base of each molecule (two bits per base). The next 2$M$ bits are placed in the second position of each molecule, next to the index information. The rest of the bits are placed in a zig-zag fashion, as shown in Figure~\ref{fig:dnamapper}. Note that regardless of how data is mapped into the matrix, the redundancy symbols are created by the Reed-Solomon encoder for each row after the mapping is done and no remapping is performed on such symbols. Once the redundancy symbols are created, every symbol in the matrix is encoded into DNA bases and each column is synthesized into a molecule.

\subsection{Determining Priority of Bits}
Priority of bits within a file can be determined for data types that have a notion and a metric of quality, such as images and videos. Previous work has proposed techniques for classifying bits into reliability classes based on the amount of damage that is caused by corrupting a bit in a given class for progressively encoded images~\cite{guo:approximate} as well as for H.264 videos~\cite{jevdjic:approximate}. Different classes of bits are then stored separately according to their reliability needs. These techniques require  additional metadata about the placement of the bits to be encoded into the storage substrate so that the files can be reconstructed from bits stored in different locations, and such metadata has to be stored in the most reliable locations~\cite{guo:approximate, jevdjic:approximate}. 

While the techniques proposed by prior work are a great use case for our system, they are not open-sourced, and they are quite complex, so we leave their integration for future work. In this work, we propose a very simple and effective technique for determining the importance of bits for standard JPEG images that does not require any additional metadata to be stored. The basic idea comes from two simple observations: 
\begin{itemize}
  \item Pixels in JPEG images are grouped into small encoding units. Each unit is placed in a JPEG file such that each depends only on the previously encoded units.
  \item JPEG uses highly efficient, but error-prone entropy-coding. Corrupting a bit in a JPEG file may confuse the entropy coder in such a way that precludes the decoding of the subsequent bits.
\end{itemize}

\begin{figure}
	\centering
	\includegraphics[width=\columnwidth]{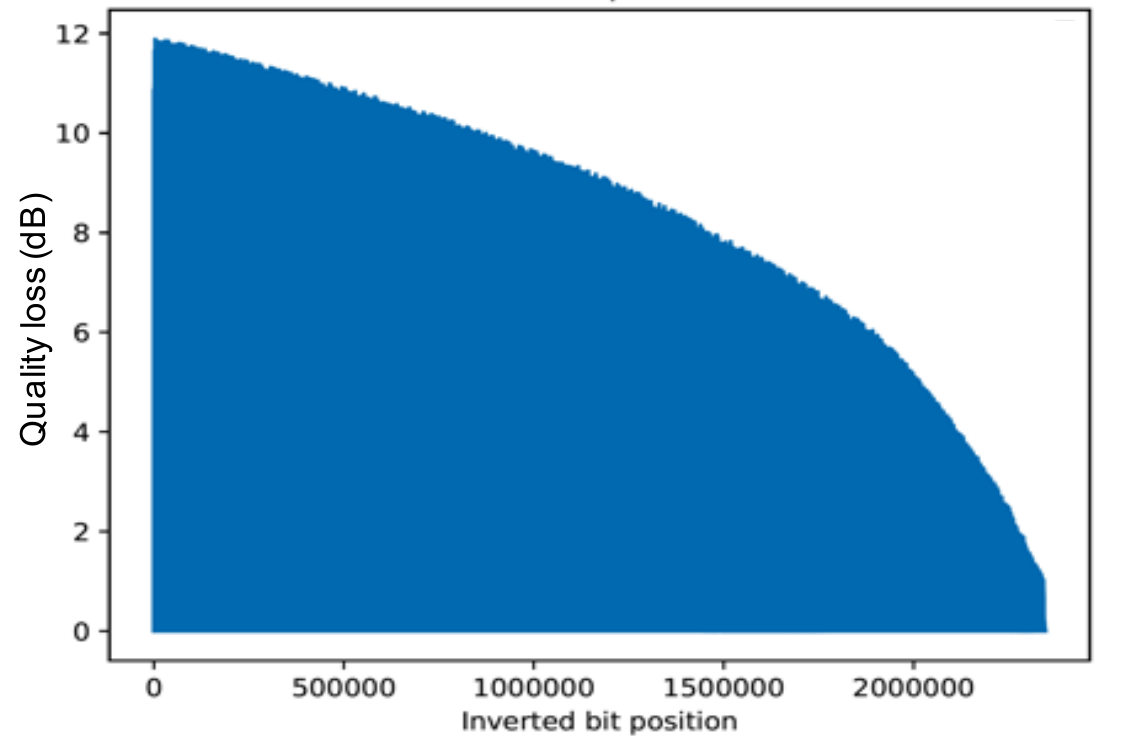}
	\caption{PSNR quality loss in dB as a function of the corrupted bit position}
	\label{fig:marbles}
\end{figure}

These two observations imply that bits that come earlier in the file tend to need more reliable storage compared to the bits that come later. To validate this observation, we profile a JPEG image by flipping one bit at a time, decoding the resulting image and measuring the quality loss with respect to the original image. Figure~\ref{fig:marbles} shows the PSNR quality loss in decibels based on the position of the bit in the file. We can see that the maximum loss is incurred by corrupting the bits at the beginning, and the minimum loss for doing so with the bits at the end of the file. Based on these observations, we simply prioritize bits based on their location in the image file.

\subsection{Using DNAMapper}
In contrast to unequal error correction, DnaMapper does not require the knowledge of the exact magnitude of the reliability skew; given the ranking of data bits by their reliability needs, DnaMapper only requires  the ranking of DNA storage locations by reliability (which can be easily established and does not change with the technology) to optimally map data to DNA. DnaMapper can be used to achieve graceful quality degradation in the presence of high error rates, or to implement the concept of approximate storage. DnaMapper retrieves data (e.g., images or videos) using the minimum amount of resources required for reconstructing the data at the desired quality level. Similarly, if the invested resources are not sufficient to fully recover the data or if the retrieving process introduces more errors than what can be corrected, DnaMapper is still capable of retrieving useful data of sufficiently high quality. In other words, as the noise level increases, the quality of DnaMapper-stored data gracefully degrades and still provide useful data, but of gradually lower quality.

\section{Methodology}
\label{sec:methodology}

\subsection{Simulation}
To evaluate the proposed techniques with realistic system parameters, we perform our evaluation using simulation. Not being financially limited  by the cost of synthesis, we assume longer DNA strands of up to 750 bases, and a set of large files of variable sizes. We put together a group of 10 images of different resolutions and qualities, whose size varies between 5KB and 1.5MB.  All images are encrypted, and the total size of all files is 8.7MB. We encode all the files into the same encoding unit (matrix) to demonstrate how files of different sizes can be mixed in a practical manner, while being compatible with both Gini and DnaMapper. For the sake of completeness, an additional file containing the names and sizes of all files has been encoded together with the files and acts as a directory, which in case of DnaMapper was given the highest priority.

\subsubsection{Storage Architecture.}
Out of 750 bases in a DNA strand, 40 bases are reserved for a pair of access primers, and 8 bases (equivalent to 16 bits) for the ordering information, and the remaining 656 bases are used for data. We use Reed-Solomon codes with 16-bit symbols and ${2}^{16}$-1 = 65535 symbols per codeword, as in the prominent DNA storage demonstrations~\cite{organick:random}. Each DNA molecule can hold exactly 82 Reed-Solomon symbols, and therefore, our Reed-Solomon matrix has 82 rows and holds up to 10.5MB of data and redundancy. We use 18.4\%  of symbols in each codeword for redundancy, leaving us with 8.7MB of pure data per unit of encoding. 

We evaluate three techniques:1) the baseline architecture~\cite{organick:random}, which is unaware of the skew, 2) Gini, which interleaves the codewords, and 3) the priority-based mapping scheme described in Section~\ref{sec:dnamapper}, where the priority of a bit is approximated by its position in the image file. In case of the priority mapping, we run into an interesting problem of how to rank the bits by reliability when we have multiple files of different sizes. Among a few heuristics we tried, the following one turned out to be the fairest and best performing: given N classes of reliability (i.e., N rows in the matrix), we give each file a fraction of storage in each reliability class in proportion to the file size. This means that the high-order bits of all files will be stored in the strongest reliability class, and the low-order bits of all files will be in the weakest class. Under this heuristic, we noticed that the presence of errors affects all files similarly in terms of the image quality loss, regardless of the image size. The only exception is the directory file, which was given the highest priority for all of its bits.

\subsubsection{Data retrieval and decoding.}
We simulate the retrieval process by creating a number of copies of each encoded DNA string and injecting insertion, deletion, and substitution errors. The error rates of modern sequencing methods vary greatly, from around 1\%~\cite{organick:random} for high-end Illumina next-generation sequencers (NGS), to 12-15\%~\cite{duda:fundamental} for low-cost nanopore-based sequencers~\cite{jain:nanopore}. We simulate a spectrum of error rates to account for a variety of sequencing methods. We also vary the average coverage between 3 and 20 by creating an appropriate number of noisy copies. The decoding process starts with data clustering. Fortunately, since we know the source strand for each noisy copy, our data is perfectly clustered, which allows us to eliminate the effects of imperfect clustering algorithms.  For consensus finding, we use the two-sided approach~\cite{organick:random}, as the other notable algorithm ~\cite{sabary:reconstruction} does not always produce the output of desired length.

To simulate different reading costs, we vary the coverage by generating a large pool of noisy strands for each DNA string. We start at a low coverage, and progressively add more strands from the pool. For every coverage point, we decode the reconstructed strands back into binary data, reassemble them into one piece, correct the errors, recombine the bits into individual files based on the directory information, decrypt every file, and finally evaluate the quality of the resulting images. We repeat this process 50 times for each data point and report the averages.

\subsection{Wetlab validation}
To validate our tool-chains end-to-end, we performed wetlab experiments in which we synthesize two small images in DNA organized with various organizations (baseline, Gini, DnaMapper), and later retrieve, sequence using NGS (at 0.3\% error rate), and successfully decode all of them. Figure~\ref{fig:gigi} (left) shows one of the successfully decoded photos. Note that the software toolchains we use for data encoding and decoding are identical for both simulated and wetlab experiments. The only difference is that when simulating, we provide fake input data with the desired magnitude of errors instead of the data that would come from sequencing. We present results only for simulation since the impact of the proposed techniques on ultra-low error rates with NGS is negligible.

\section{Evaluation}
\label{sec:evaluation}

\subsection{Gini}
\begin{figure}
	\includegraphics[width=\columnwidth]{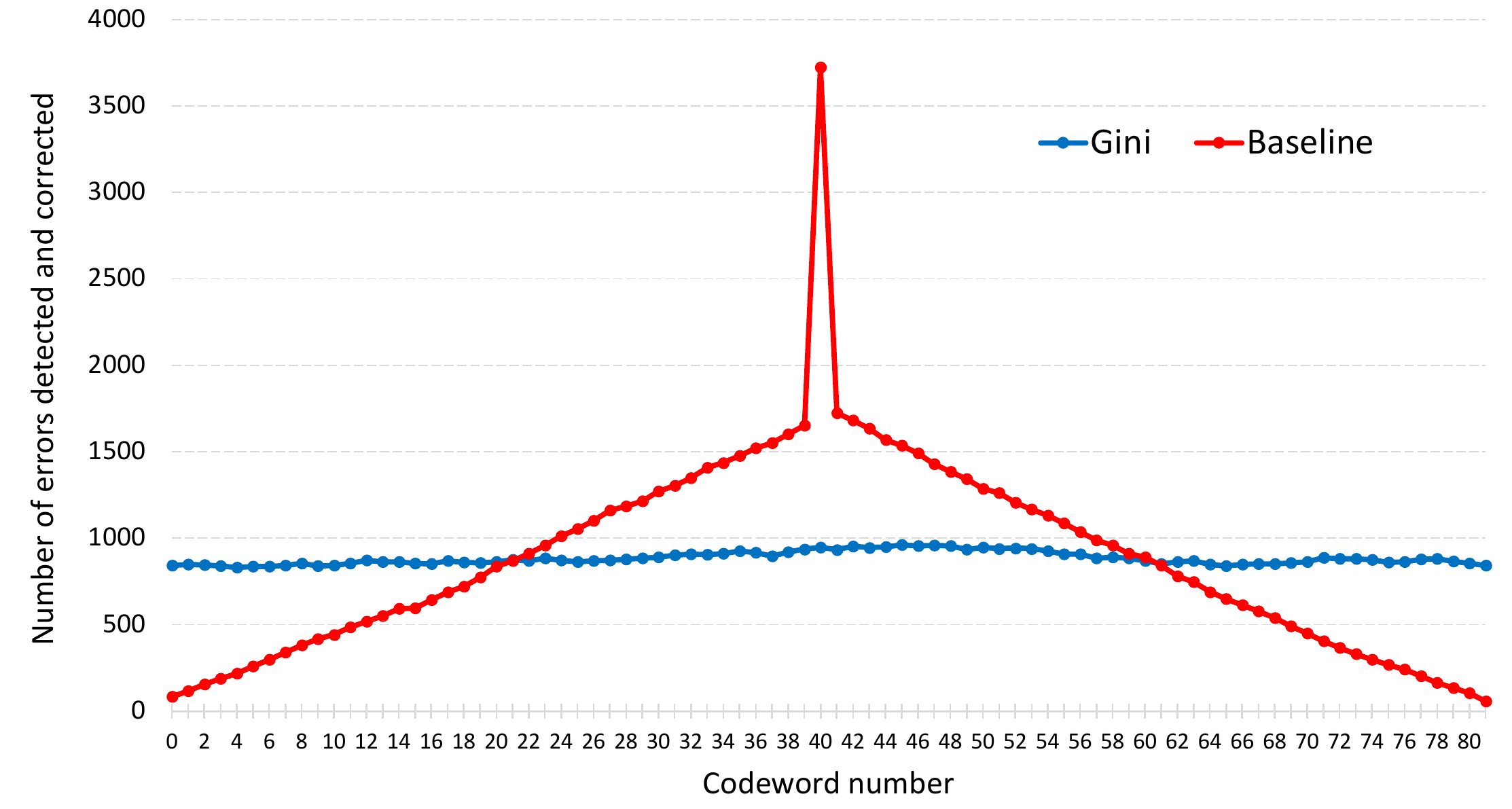}
	\caption{Positional distribution of errors per codeword in baseline (red) and Gini (blue) at the error rate of 9\% and sequencing coverage of 20.}
	\label{fig:gini_skew}
\end{figure}

Figure~\ref{fig:gini_skew} shows the number of errors each codeword receives in case of using the baseline, where each codeword is a row in the matrix, and Gini, where each codeword is diagonally striped across the matrix. The experiment is done at the error rate of 9\% and sequencing coverage of 20. We can see that for the baseline, rows that are closer to the ends experience significantly fewer errors at the expense of the rows in the middle, with a prominent peak in the middle. In contrast, Gini's interleaving of codewords across both rows and columns ensures that every codeword experiences a similar number of errors, effectively flattening the curve and removing the bias. Note that the surface under the curves, which corresponds to the total number of errors, is the same.

\begin{figure}
	\includegraphics[width=\columnwidth]{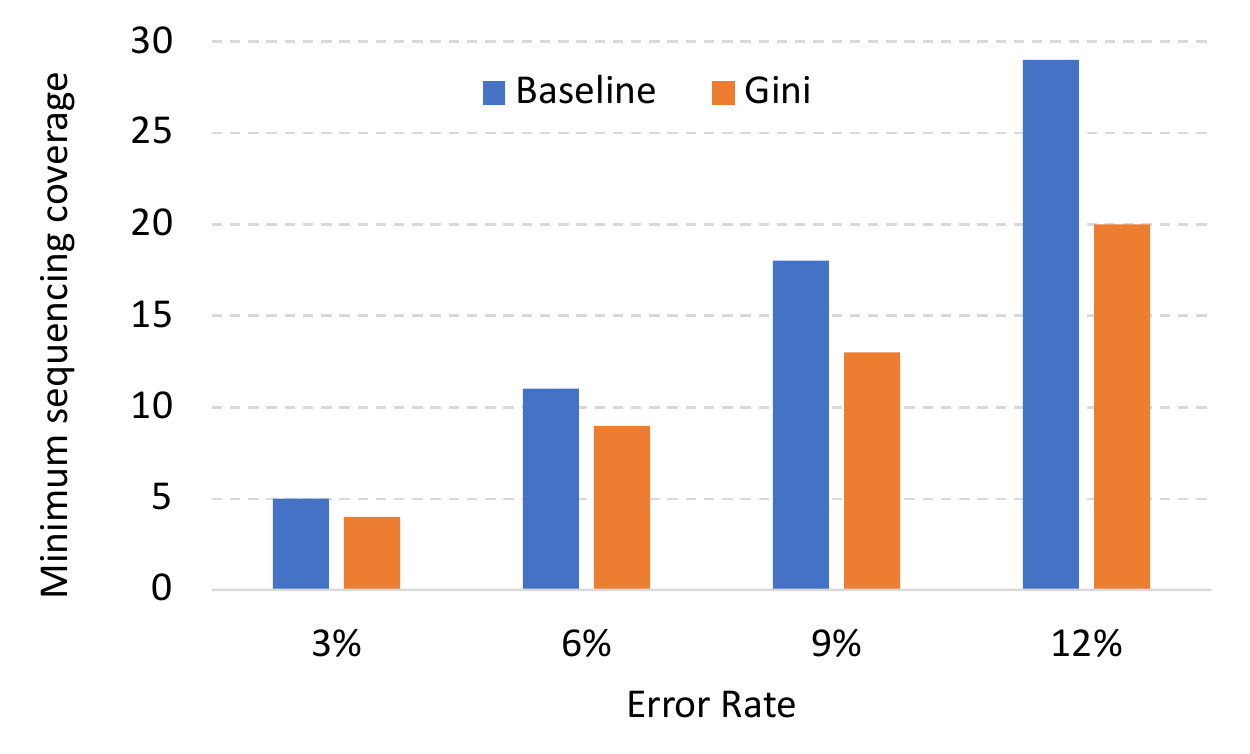}
	\caption{Minimum sequencing coverage required for error-free decoding as a function of error rate.}
	\label{fig:gini_mincov}
\end{figure}

    Because Gini removes the bias, it requires lower sequencing coverage to retrieve the data exactly, without any errors, compared to the baseline which must provision for the worst case. Figure~\ref{fig:gini_mincov} shows the minimum sequencing coverage (the lower the better) required for error-free decoding as a function of error rate, for the baseline and Gini. As we can see, Gini can reduce the required sequencing coverage, and therefore the reading cost, by 20\% for small error rates, and up to 30\% for higher error rates. Similarly, if the same sequencing coverage is used for both Gini and baseline, Gini would have significantly higher chances of exact error-free decoding, increasing the reliability of the system.

\begin{figure}
	\includegraphics[width=\columnwidth]{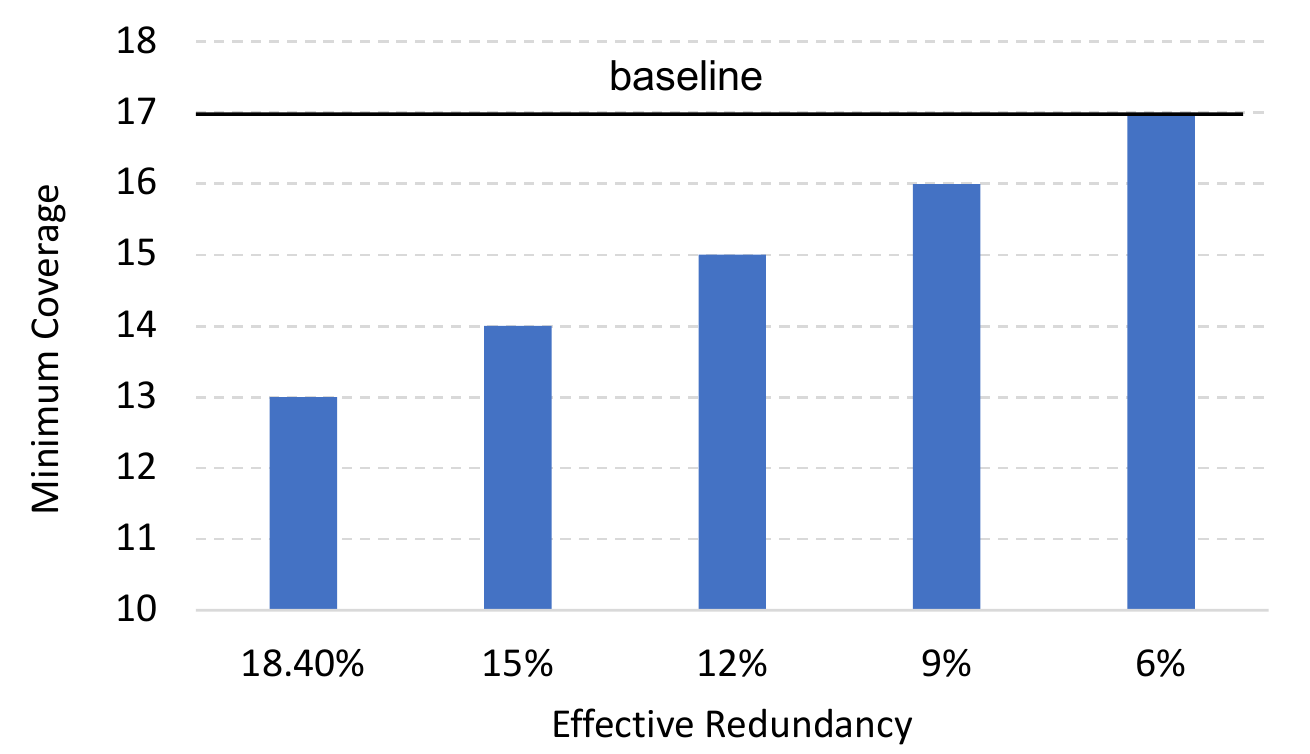}
	\caption{Minimum sequencing coverage required for error-free decoding as a function of effective redundancy. The error rate is fixed at 9\%.}
	\label{fig:gini_writes}
\end{figure}

To evaluate Gini's potential for savings in synthesis cost, we fix the error rate to 9\% and gradually reduce the amount of Gini's error correction resources until Gini matches the coverage of the baseline at that error rate (17). We simulate the reduction in error-correction resources by introducing erasures in a controllable manner, so that the effective redundancy is reduced. Figure~\ref{fig:gini_writes} shows that Gini's redundancy can be reduced from 18.4\% to only 6\% while matching the coverage requirements of the baseline, which is a 67\% reduction in redundancy and 12.5\% reduction in the entire synthesis cost. 

\subsection{Data Mapping}

\begin{figure*}
	\includegraphics[width=1\textwidth]{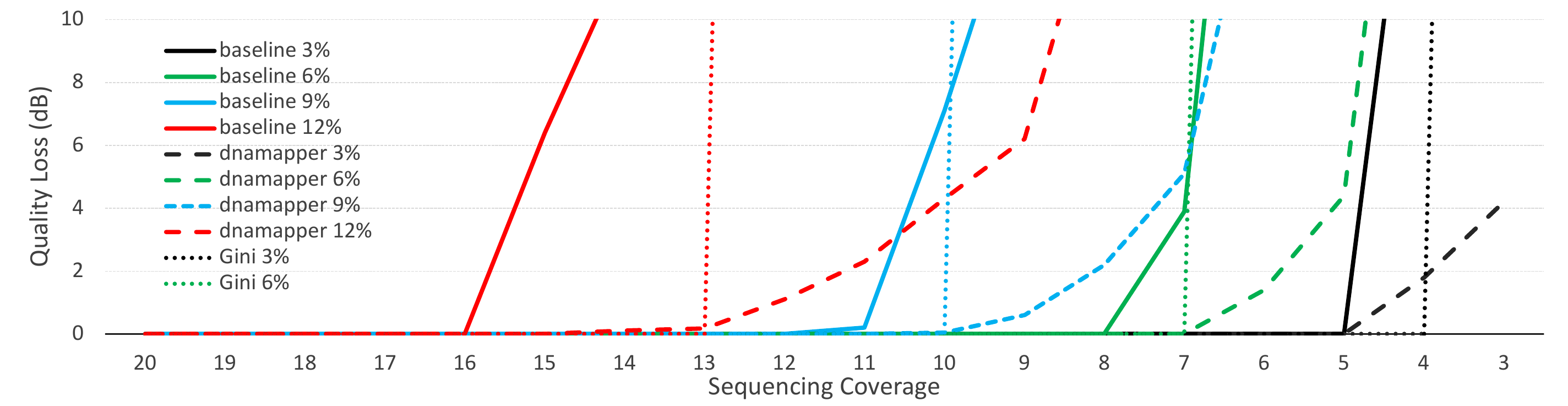}
	\caption{Quality loss of retrieved images as a function of coverage at various error rates. Full lines denote the baseline data mapping, and dashed lines denote DnaMapper, and dotted line denote Gini.}
	\label{fig:priority}
\end{figure*}

\begin{figure}
	\includegraphics[width=\columnwidth]{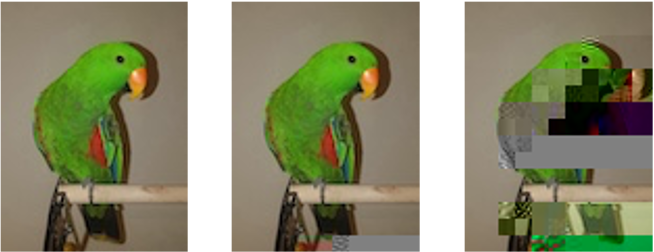}
	\caption{Original image (left), 1.2dB loss (middle), and 7.1dB loss (right).}
	\label{fig:gigi}
\end{figure}

Figure~\ref{fig:priority} shows the quality loss in decibels for images retrieved from the simulated DNA storage system in case of the baseline data mapping, and the proposed DnaMapper scheme, as well as Gini, while varying the coverage from 3 to 20. For very low error rates, all systems can successfully decode the files at any coverage. However, as we increase the error rate, we can see that the image quality degradation experienced by the baseline system is increasing sharply as we reduce coverage. In contrast, DnaMapper loses quality very gradually as coverage decreases.  For example, for error rate of 12\% and coverage of 13, the baseline experiences catastrophic data loss such that the image cannot be decoded. In contrast, DnaMapper only loses 0.3dB in quality, which is not even noticeable (up to 1dB loss is considered unnoticeable ~\cite{guo:approximate}; as a reference, Figure~\ref{fig:gigi} shows an example of decoded images. The image on the left was successfully decoded with no errors. The image in the middle suffered 1.2dB loss, while the last image suffered 7.1dB loss.) The gap between DnaMapper and the baseline increases with the error rate, leading to 20-50\% reduction in reading cost for the same quality target. As in the case of Gini, an identical analysis could be performed for redundancy savings when coverage is kept constant, but we omit that for brevity.
 
 It is important to note that Gini (dotted lines) reduces the coverage needed for error-free decoding by flattening the error curve, and as long as the number of errors is below the threshold that the codewords can handle, every codeword will be decoded without a single uncorrected error. However, the moment the coverage drops beyond the threshold, all of a sudden all codewords  fail to decode at the same time, as this threshold is crossed in all codewords simultaneously due to Gini's interleaving. As a result, we can see that Gini can occasionally perform even worse than the baseline, which in the high-error regime can at least decode some rows that are far from the middle. 

In contrast to Gini, DnaMapper will ensure \textit{graceful quality degradation} as the level of noise increases, which is particularly interesting in scenarios when the noise levels cannot be predicted. It also allows us to trade quality for cost in a controllable manner. Compared to Gini, DnaMapper tends to suffer minor quality loss in the medium noise range, as it does not flatten the curve. For example, at coverage of 14 for error rate of 12\%, there is a quality degradation of 0.03dB, and 0.1dB at the coverage of 13; in contrast, Gini results in error-free retrieval at coverage of 14, but at coverage of 13, Gini's output is not decodable.

Note that we use PSNR as an image quality metric because it is an objective metric, known to reserchers even outside of the media-processing community. However, we believe that a subjective quality metric that includes the user perception would be more relevant. The study of such metrics is beyond the scope of this work.

\subsection{Evaluating the Bit Ranking Method}
\label{sec:oracle}
\begin{figure}
\begin{center}
	\includegraphics[width=\columnwidth]{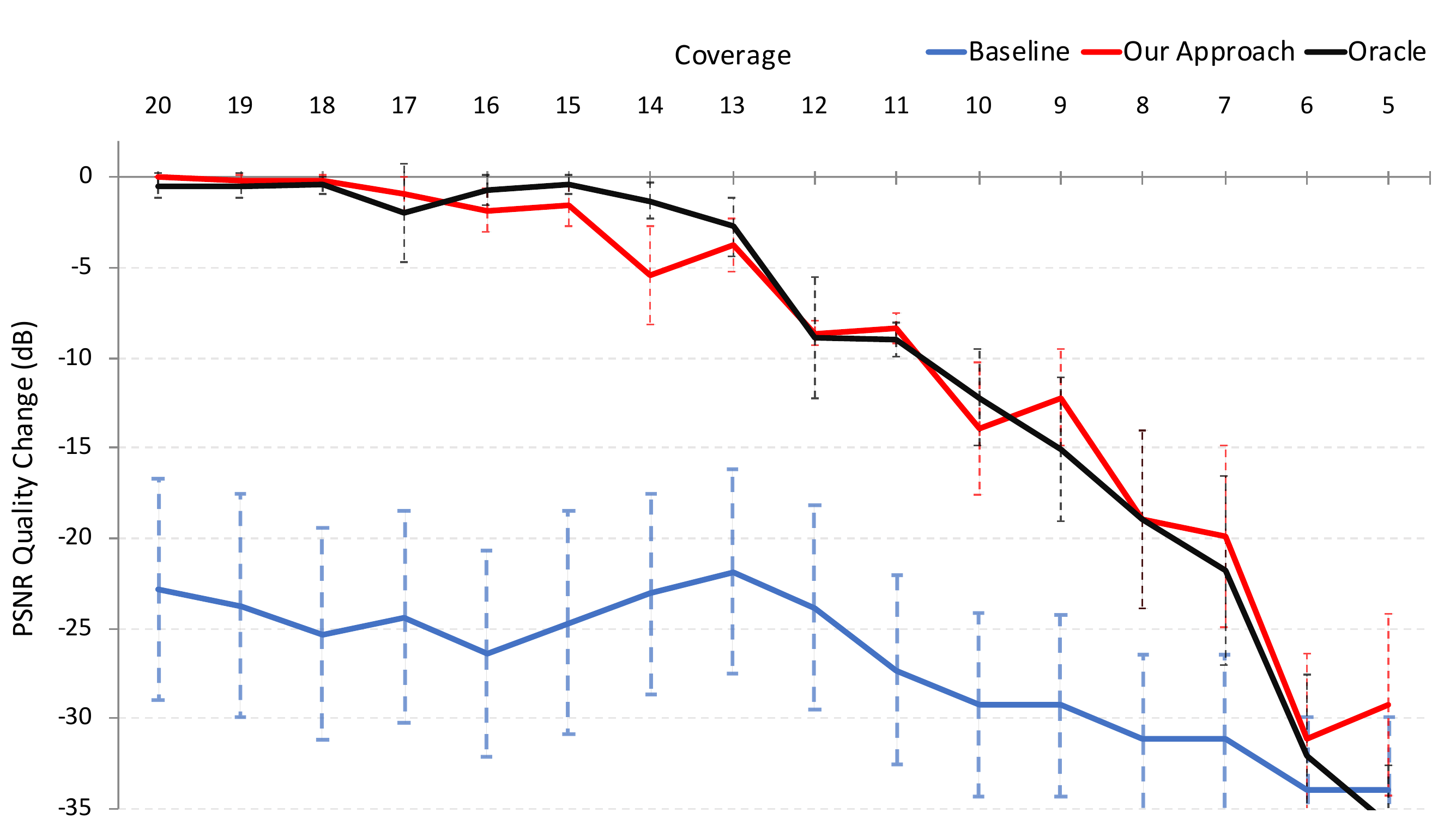}
	\caption{Comparison of the proposed reliability ranking method with the ideal oracle ranking.}
	\label{fig:oracle}
\end{center}
\end{figure}

Figure~\ref{fig:oracle} compares our simple bit ranking heuristic against an oracle ranking. To determine the oracle ranking, we use the following brute force method: we corrupt an image file by flipping one bit at a time and evaluate the quality loss in decibels (dB) using the peak signal-to-noise ratio (PSNR) as a quality metric. We then sort all the bits according to the quality loss, such that the bits that cause higher quality loss upon flips are ranked higher and placed in positions of higher reliability. We benchmark the oracle approach using a medium size image file (300KB), for which it was feasible to compute an oracle ranking. In this experiment, we don't use any error correction.

Note that the oracle method does not perform visibly better compared to our method, despite relying on a computationally expensive exhaustive search and requiring an unacceptable storage overhead to store the rankings. Note that errors in DNA storage are not independent from each other, nor their impact on the quality of the decoded image is independent.  For example, the second of two consecutive errors in neighboring bits in an image is much less likely to affect the quality compared to the first one. This is something that our oracle cannot capture. The presented “oracle” is thus the best approximation of the actual oracle that was possible to evaluate. However, we believe that more sophisticated bit ranking methods can be developed for various types of data, but these are out of scope for this paper. Also note that our proof-of-concept ranking method incurs zero storage overhead and is extremely simple, and allows for end-to-end encrypted files to be stored, as the bit ranking is determined without looking at the content of the images. 

\section{Discussion and Related Work}
\label{sec:disc}

\textbf{Impact on locality}. This work introduces no new locality trade-offs by shuffling contiguous data bits across different molecules. The reason is that an encoding unit must be encoded, fetched, and decoded together with all (or most, erasures allowed) data present, and that’s regardless of whether or not data is internally shuffled. Locality across encoding units is also unaffected by our techniques.

\textbf{Breakdown or Error Types}
A large-scale study has found that in a typical data retrieval workflow that uses NGS, about 25-30\% of errors are indels (insertions and deletions), the rest being substitutions~\cite{organick:random}. The same study has found that over 60\% of errors are indels in nanopore-based workflows. Both workflows assume conventional DNA synthesis, where most of the time and resources is spent on ensuring that every base is synthesized exactly once. The emerging enzymatic synthesis technology ~\cite{lee:enzymatic} relaxes this rule, which dramatically inflates the number of indels at the moment of synthesis (e.g., ACGT can be synthesized as AAACTT), which we expect to further exacerbate the skew problem.

\textbf{Realistic Error Rates} DNA Sequencing methods have been evolving in the direction of higher noise for the past 45 years. The oldest sequencing method used today is Sanger sequencing from the 1970s, which is also the most accurate today, but impractical as it correctly reads only DNA pools dominated by a single DNA strand. Next-Generation Sequencing (NGS) appeared two decades ago with lower accuracy, but much higher throughput. Nanopore-based methods that evolved recently introduce orders of magnitude higher errors, but bring other benefits such low-cost real-time sequencing. DNA-based storage at the right scale still needs many orders of magnitude improvement in sequencing cost, latency, throughput, and environmental cost, and it's quite possible that error rates in such methods will be over 30\%~\cite{sabary:reconstruction}, or even higher if the enzymatic synthesis~\cite{lee:enzymatic} is used.

\textbf{Relationship to Approximate Storage/Memory}
 Similarly to prior work~\cite{guo:approximate, jevdjic:approximate}, DNAMapper also uses heuristics to rank bits by reliability for a given data type. However, all prior work on approximate storage/memories assumes uniformly reliable storage medium that is deliberately engineered into reliability classes (e.g., by selectively reducing DRAM refresh rate~\cite{flikker, dram1, dram2}, adding different amounts of redundancy in Flash/PCM~\cite{guo:approximate, jevdjic:approximate}, etc.) to match various reliability needs in the data. In contrast, DNA is the first substrate that has an intrinsic and dynamically changing reliability skew. DNAMapper is the first technique that stores data with varying reliability needs into a substrate with varying reliability properties.

\section{Conclusion}
\label{sec:conclusion} 
In this paper we made a novel observation that the probability of successful recovery of a given bit from any type of a DNA-based storage system highly depends on its physical location within the DNA molecule. We showed that the reliability skew is fundamental to all DNA storage systems, and that it  leads to highly inefficient use of error-correction resources and higher synthesis and sequencing costs. We proposed two approaches to address the problem. The first approach, Gini, distributes the errors across error-correction codewords in a way that equalizes the impact of errors across many codewords, without increasing the size of the encoding unit. This approach effectively removes the positional bias and reduces the associated costs. The second approach, DnaMapper, seeks to leverage the bias and relies on application-aware mapping of data onto DNA molecules such that data that requires higher reliability is stored in more reliable locations, whereas data that needs lower reliability is stored in less reliable parts of DNA molecules, reducing the cost of sequencing and providing graceful degradation. All proposed mechanisms involve no storage overhead, can be directly integrated into any DNA storage pipeline.

\begin{acks}
The authors would like to sincerely thank Lara Dolecek, Cyrus Rashtchian, and Sergey Yekhanin for many useful discussions that took place in the earlier stages of this work, as well as Cheng-Kai Lim for his immense help with wet-lab experiments. The authors also thank our shepherd, Hung-Wei Tseng, as well as all other reviewers for their valuable feedback. This work was partially funded by the Advanced Research and Technology Innovation Centre (ARTIC) at the National University of Singapore, project FCT-RP1.
\end{acks}

\bibliographystyle{ACM-Reference-Format}
\bibliography{references}

\end{document}